\documentclass[xaps,twocolumn]{revtex4}

\usepackage{amsmath}  % needed for \tfrac, \bmatrix, etc.
\usepackage{amsfonts} % needed for bold Greek, Fraktur, and blackboard bold
\usepackage{graphicx, subfigure} % needed for figures

\usepackage{color}
\definecolor{b}{rgb}{0,0,1}

\definecolor{m}{rgb}{1,0,1}

\definecolor{r}{rgb}{1,0,0}

\definecolor{g}{rgb}{0,1,0.2}

\begin{document}

\title{Silo outflow of soft frictionless spheres}

\author{Ahmed Ashour$^{1,2}$, Torsten Trittel$^1$, Tam\'as B\"orzs\"onyi$^3$, and Ralf Stannarius$^1$}
\affiliation{$^1$Institute for Experimental Physics, Otto von Guericke University, Magdeburg, Germany\\
$^2$Faculty of Engineering and Technology, Future University, end of 90 St., New Cairo, Egypt\\
$^3$ Institute for Solid State Physics and Optics, Wigner Research Center for Physics,
Hungarian Academy of Sciences, %P.O. Box 49, H-1525
Budapest, Hungary}

\date{\today}

\begin{abstract}
Outflow of granular materials from silos is a remarkably complex physical phenomenon that
has been extensively studied with simple objects like monodisperse hard disks in two dimensions (2D) and hard spheres in 2D and 3D. For those materials, empirical equations were found that describe the discharge characteristics.
Softness adds qualitatively new features to the dynamics and to the character of the flow. We report a study of the outflow of soft, practically frictionless hydrogel spheres from a quasi-2D bin. Prominent features are intermittent clogs, peculiar flow fields in the container and a pronounced dependence of the flow rate and clogging statistics on the container fill height. The latter is a consequence of the ineffectiveness of Janssen's law: the pressure at the bottom of a bin containing hydrogel spheres grows linearly with the fill height.
\end{abstract}

\maketitle

%\section{Introduction}

Granular materials are being handled by humans since ancient times in agriculture, mining, building and
other activities, and intuitively we are all familiar with some of their unusual properties in typical everyday-life situations, when compacting the granular content of a bag by shaking, pouring salt or sugar from a muffineer, or dealing with clogging of the outflow from storage containers.
Nevertheless, the comprehensive physical description is still incomplete today
\cite{Jaeger1996,deGennes1999,Campbell2006,Carpinlioglu2016}.
The flow of grains, even in the simplest form of hard monodisperse spheres, has retained some mysteries.

Hoppers are storages that exploit the ability of granular matter to flow liquid-like through sufficiently large outlets. Frequent problems with such storage devices involve, in particular,
the formation of clogs that interrupt flow, and the redistribution of weight to the side walls by
force chains, which may cause structural collapses. For hard spheres and, in 2D, hard disks,
previous research activities brought important insights and quantitative laws that describe the
discharge dynamics and static conditions reasonably well.
Through sufficiently large orifices, such grains flow rather
continuously. Quantitative predictions for flow rates were proposed
(e. g. \cite{Franklin1955,Beverloo1961,Neddermann1982,Zuriguel2007} and references therein). When the outlet diameter is
smaller than about 5 particle diameters \cite{To2001,To2005,Zuriguel2005,Thomas2015,Mort2015}, clogs form after some time at the orifice and block further
outflow, mostly unwanted.
Spontaneous arch formation~\cite{Tang2011,Zuriguel2014_1}, the preceding kinetics
\cite{Rubio-largo2015}, as well as the inherent force distributions~\cite{Hidalgo2013,Vivanco2012} have been analyzed in the literature.
In order to study the geometry and statics of such clogs, often 2D container geometries are chosen.

Identical hard spheres are idealized special cases: In practice one often deals with polydispersity, non-spherical grains, and/or elasticity of the material. Even small modifications of these features can have dramatic qualitative consequences
\cite{Ashour2017}. However, studies of elastic particles, for example, are rather scarce.
Experiments with dense air bubble arrangements in an 'inverse' 2D Hele-Shaw silo model have been reported by Bertho et al. \cite{Bertho2006}. In this extreme case of incompressible but highly elastic constituents, outflow rates $Q$
of bubbles with diameter $D$ through an orifice with width $W$ were found to obey a relation $Q\propto g^{1.5}(W/D-k)^{0.5}$, where $g$ is the gravitational acceleration, and $k\approx 2/3$ is a constant empirically found
in the experiment. The high deformability allowed passage of bubbles even when $W<D$, no clogging was reported.

Recently, oily emulsions were studied in a 2D hopper geometry \cite{Weeks2017,Weeks2017b}. These ensembles of polydisperse oil droplets also have many similarities to granular material, but combined with shape elasticity.
The authors reported considerable differences in the dynamic behavior, particularly, the critical ratio of orifice size and particle size is much smaller than for rigid grains, and non-permanent, intermittent clogs were observed.
Similar intermittent clogs are found with hard particles in a vibrated hopper~\cite{Zuriguel2009},
or in flocks of animals passing a gate \cite{Zuriguel2014_1}.
\begin{figure}[htbp]
\includegraphics[width=0.8\columnwidth]{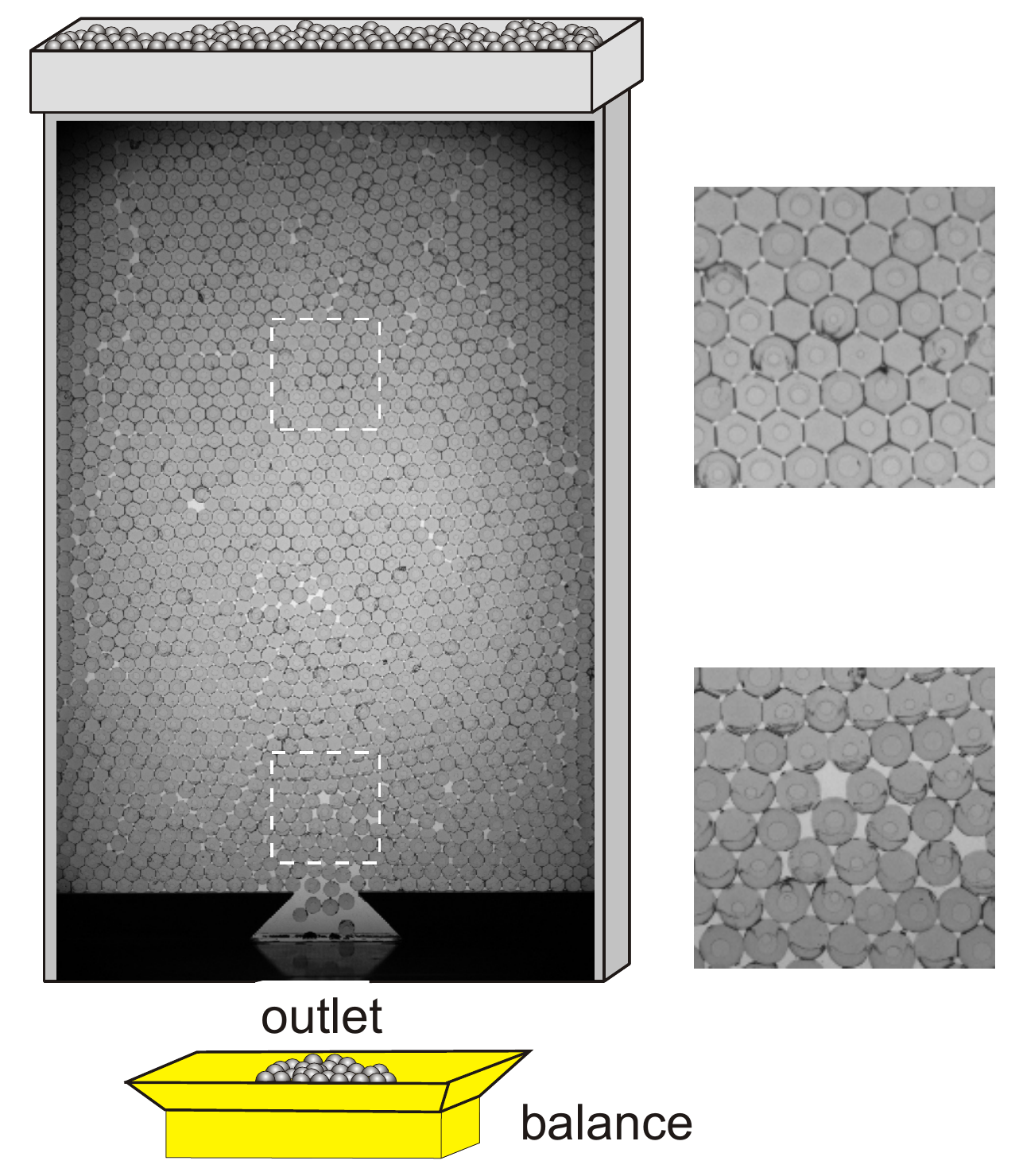}
	\caption{Experimental setup: The small clips at the right show details of the hexagonal lattice
formed by the soft spheres far from the container walls in the upper half of the bin, and directly above the outlet (positions indicated by dashed squares).
}
\label{fig:setup}
\end{figure}

Soft monodisperse hydrogel spheres \cite{Bertrand2016,Dijksman2017} are incompressible,
but deformable by
moderate pressures so that they can adopt their shapes to local static conditions in the container.
This changes the pressure profile and the discharge characteristic, as well
as the flow field inside the container, and leads to a fill-height dependent outflow.

In the present study, we investigate hydrogel spheres of $D=9.2$~mm diameter, acquired at {\em Happy Store, Nanjing}
in dry state and swelled with distilled water for at least 24 hours. Between experiments, the spheres were kept in distilled water.
The material differs in two aspects from the previously studied hard spheres, its softness and its small friction coefficient. For similar hydrogel spheres, a friction coefficient of $\mu=0.03$ has been estimated
\cite{Brodu2015}. Dijksman et al. \cite{Dijksman2017}
determined the Young modulus of hydrogel spheres to be $\approx 20$~kPa.
By measurement of Hertzian contact diameters we have determined an elastic modulus of our spheres
between 30 kPa and 50 kPa, where the outer part is slightly softer than the core.

Our experiments were performed in a quasi-2D bin with lateral dimensions of about 400~mm $\times$ 800~mm, and a thickness slightly larger than the sphere diameters (Figure~\ref{fig:setup}). The outlet
width $W$ and its position can be controlled arbitrarily. In the experiments presented here, the position of the outlet is in the center of the container bottom. The approximate content of the filled container is
$m\approx 4700$ spheres.
The container is filled with the soft spheres from the top, then the outlet
is opened and the weight of the material that has passed the outlet is continuously monitored.
We analyze the discharge characteristics, the formation of clogs and their stability, the velocity profiles in the bin, and static features.

\begin{figure}[ht]
\large a)\includegraphics[width=0.72\columnwidth]{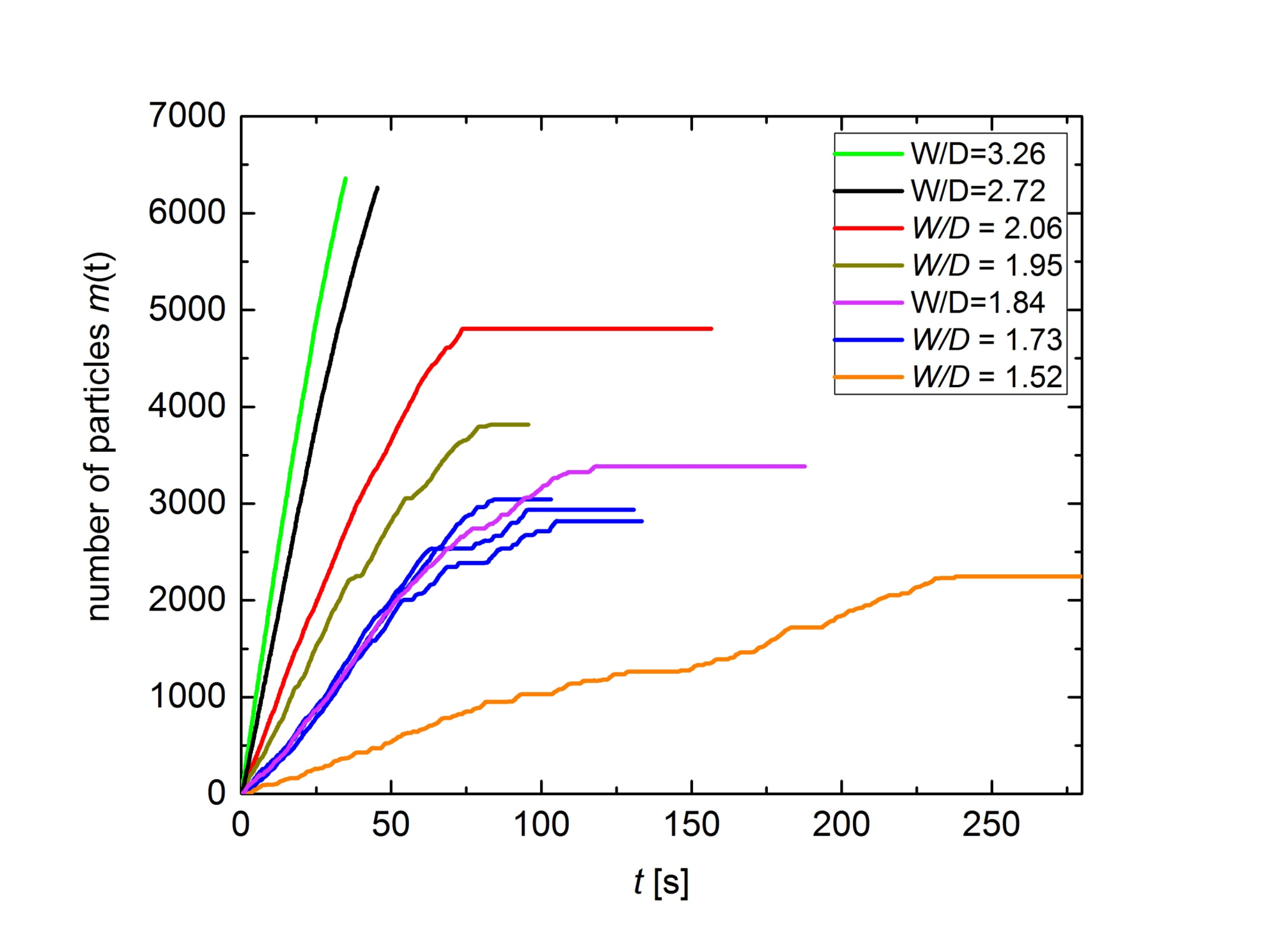}\\
b)\includegraphics[width=0.78\columnwidth]{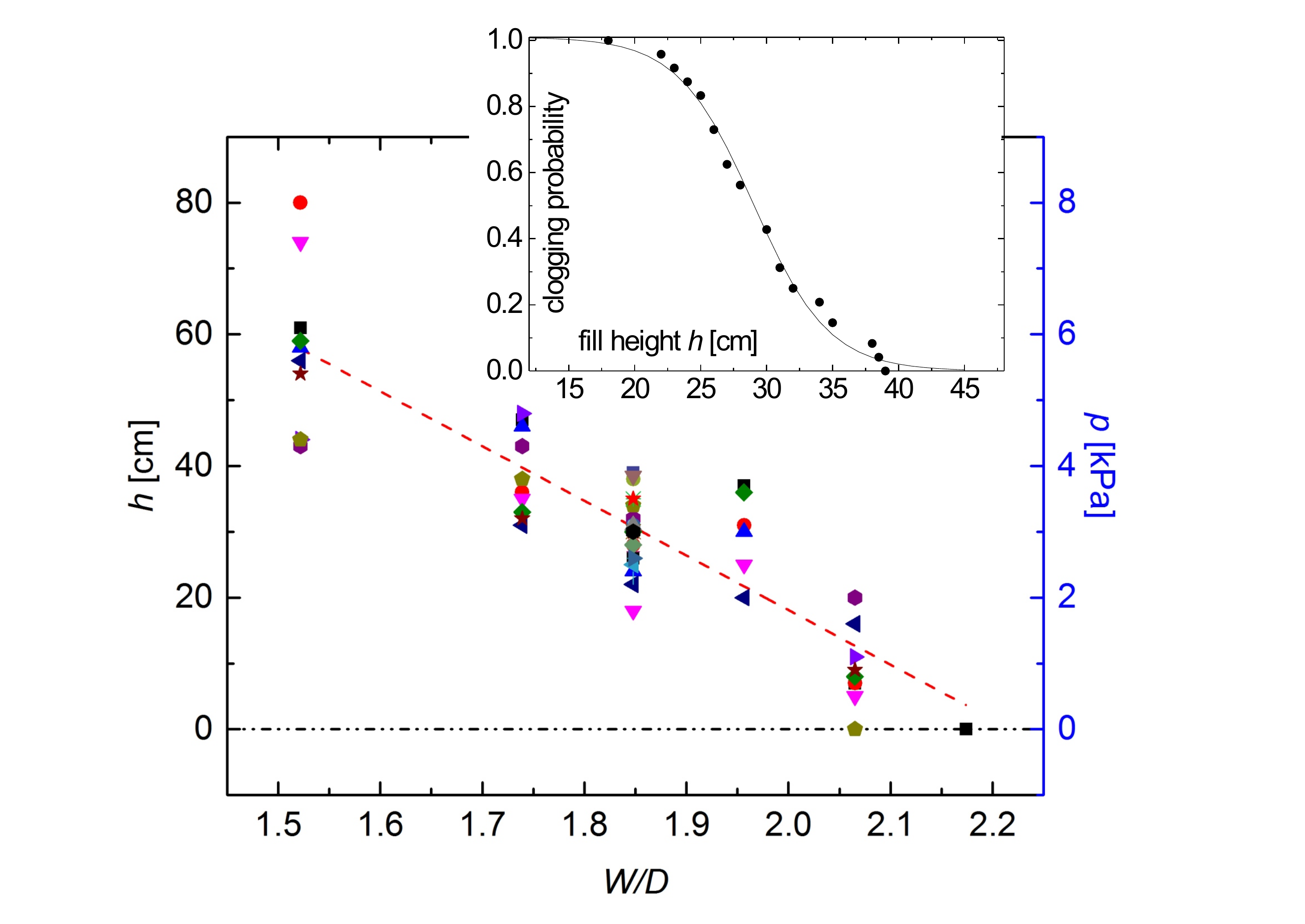}
	\caption{a) Outflow through different orifice sizes. When $W>2.2D$, no clogging is observed. With decreasing orifice size, the discharge rate decreases, the fluctuations of the
outflow velocity increase, intermittent clogs occur. At a certain fill height, the system clogs permanently (horizontal lines in the $m(t)$ discharge characteristics).
b) Remaining fill level $h$ in the permanently clogged state for different outlet sizes $W/D$.
A satisfactory linear fit yields $h_c = (2.2-W/D)\cdot 95$ cm, or, a critical pressure $p_c = (2.2-W/D)\cdot 6.6$~kPa.
For the pressure scale at the right hand side, see Fig. \ref{fig:pressure}. The inset shows exemplarily probability of
a clog at height $h$ for $W=1.85  D$. The solid hyperbolic tangent curve guides the eye.
}
\label{fig:outflow}
\end{figure}

When the outlet is closed, the monodisperse hydrogel spheres form an almost regular
lattice in the container (cf. clips in Fig.~\ref{fig:setup}). Within this lattice, the hydrogel spheres are
deformed towards a more or less hexagonal cross-sectional shape.
Thus, the packing fraction becomes larger than the maximum $\phi= \pi/\sqrt{27}\approx 0.605$ achievable for hexagonally packed hard spheres in a cell of thickness $D$. In fact, $\phi$ is roughly 0.68.

The outflow of the hydrogel spheres has a unique orifice size dependence.
When $W/D$ is larger than $\approx 2.2$, the bin empties without clogging.
This is in sharp contrast to
hard spheres where clogging already sets in when $W/D\lesssim 5$ \cite{To2001,Zuriguel2005,Thomas2013,Mort2015,Lozano2015}. Another striking difference to hard grains
is that there are only very small fluctuations of the flow velocity field inside the container for those
outlet sizes.
For outlet widths smaller than $2.2D$, the soft-sphere flow undergoes a qualitative change.
Typical discharge characteristics for $W$ between  $ 2.1D$ and $1.5 D$ are shown in Fig.~\ref{fig:outflow}a.
With decreasing ratio $W/D$, the fluctuations of the outflow rate increase.
Temporary clogs occur which can last for several seconds, they are followed by a permanent clog
at a certain fill level $h_c$. Repeating the measurement with fixed outlet size $W/D$,
the fill heights where permanent clogs occur were found to vary by about $\pm 30$~\%. The average critical height $h_c$ depends on the outlet size, a clear trend is seen in Fig.~\ref{fig:outflow}b. In first approximation, this critical fill height $h_c$ is a linear function of the orifice size.

\begin{figure}[htbp]
\includegraphics[width=0.85\columnwidth]{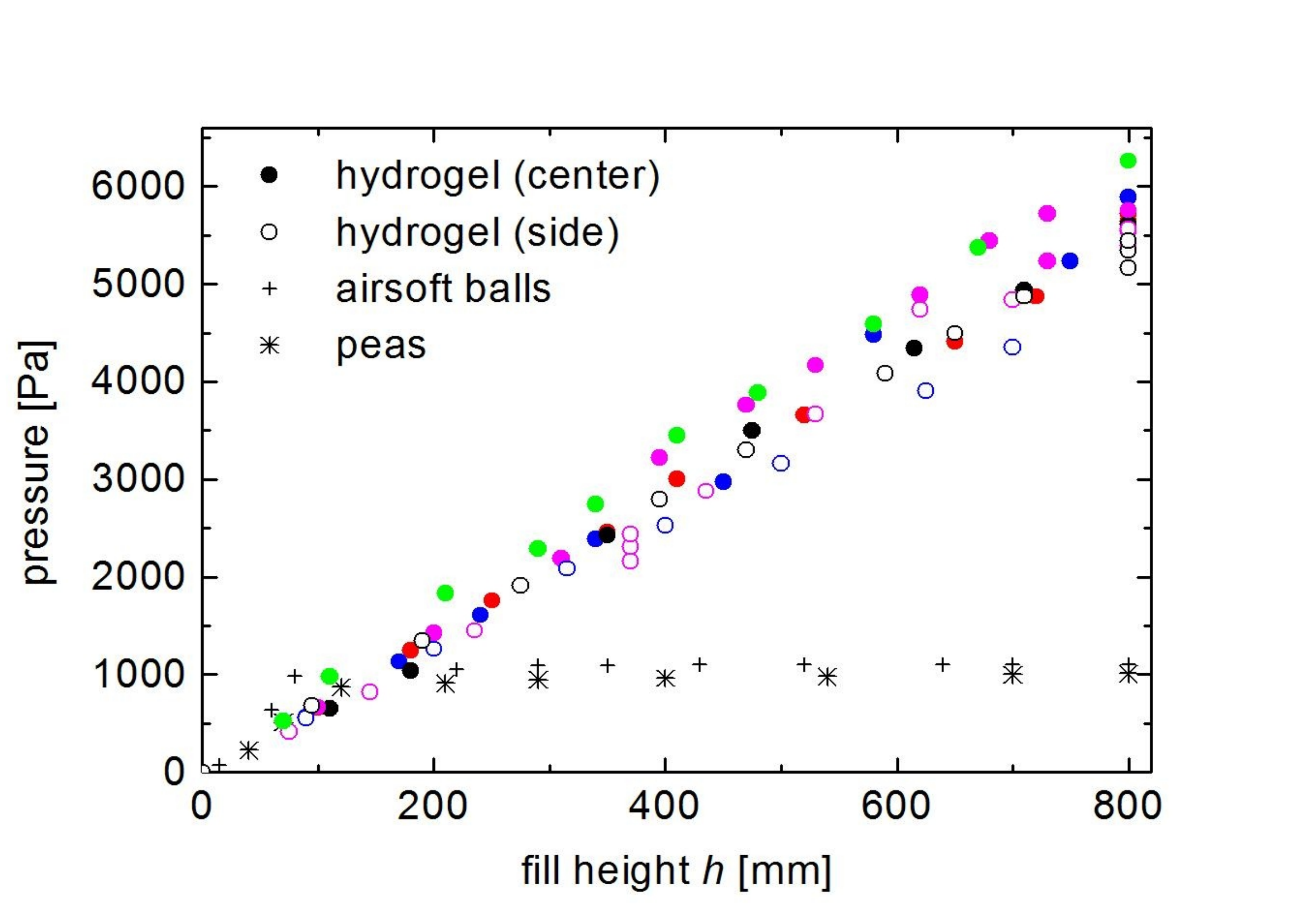}
	\caption{
Static pressure $p_0$ at the container bottom as a function of the fill height in the center of the bottom (solid symbols) and near the side of the bottom (open symbols). Different colors symbolize independent runs of the experiment. For comparison, we included the same measurement for peas (mean diameter 7.1 mm, density 1.35 g/cm$^3$, friction coefficient 0.35 determined from the angle of repose and airsoft balls (diameter 5.88 mm, density 1.05  g/cm$^3$).
There, the pressure saturates at fill heights of about 10 cm as a consequence of the Janssen effect \cite{Janssen1895,Sperl2006}.
The pressure at the bottom of the container filled with hydrogel spheres increases with fill height by $p(h)\approx h \cdot7000$~N/m$^3$.
}
\label{fig:pressure}
\end{figure}

It is intimating to
assume that such behavior is related to a hydrostatic pressure characteristics in the container. We confirmed this by measurements of the pressure $p_0$ at the bottom for different fill heights $h$. A horizontal plate 6 $\times 1$ cm$^2$  was connected to the balance and inserted at the bottom as a weight sensor, and $p_0$ was extracted from the measured weights (Fig.~\ref{fig:pressure}). There are statistical variations ($\approx 10 \%$) for each individual filling, but in general, $p_0$ was found to grow linearly with $h$. This appears to be a consequence of the low friction coefficient. No saturation was observed, in contrast to the pressure in the pea-filled and
airsoft-ball-filled containers measured for comparison. Further increasing the fill height leads to fracturing of the hydrogel beads before a noticeable deviation from the linear pressure increase would be observed.
The pressure in the center of the bottom (solid symbols) is slightly higher than near the sides (open symbols).
The density of the hydrogel is $1.03$ times that of water. The static pressure averaged over the bottom amounts to about
$\approx 68$~\% of the hydrostatic pressure in a water-filled vessel. The difference to a water column originates primarily from the packing fraction $\phi$. From the pressure data it seems reasonable to conclude that there is practically no stress relaxation by force chains towards the lateral walls.
Note that when the bin discharges, the pressure $p$ on the bottom plate drops by up to 50 \% compared to $p_0$ at same momentary fill heights.

\begin{figure}[htbp]
\includegraphics[width=0.95\columnwidth]{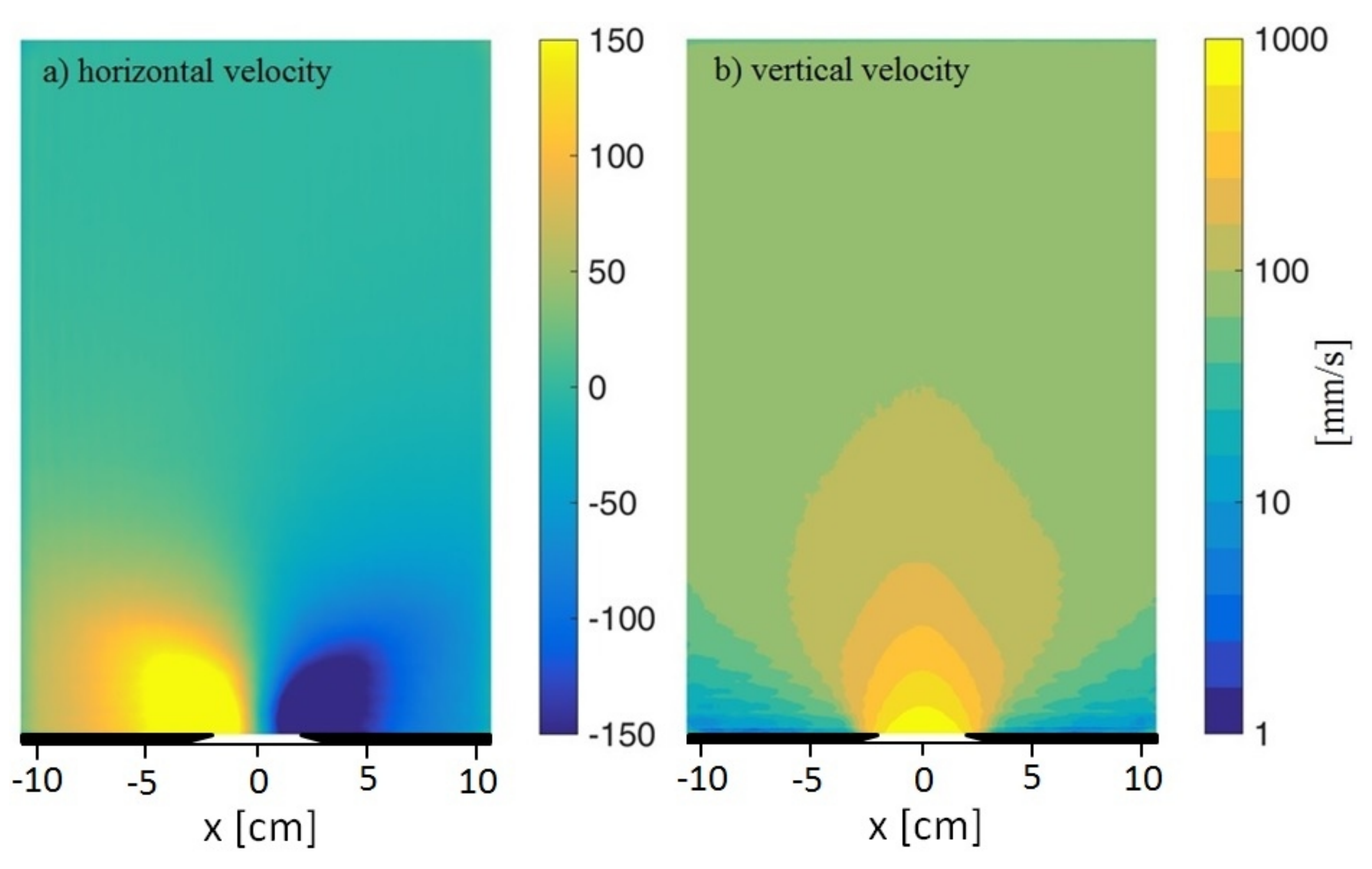}
	\caption{
Flow profiles (mm/s) at an orifice size $W=60$~mm ($W/D=6.5$): a) horizontal flow, the yellow (bright) region indicates flow to the right, the dark (blue) region to the left, green indicates zero horizontal flow; b) vertical flow. The images show a 23 cm $\times$ 34 cm region of the container above the
orifice.
}
\label{fig:flowfield}
\end{figure}

\begin{figure}[htbp]
\includegraphics[width=0.8\columnwidth]{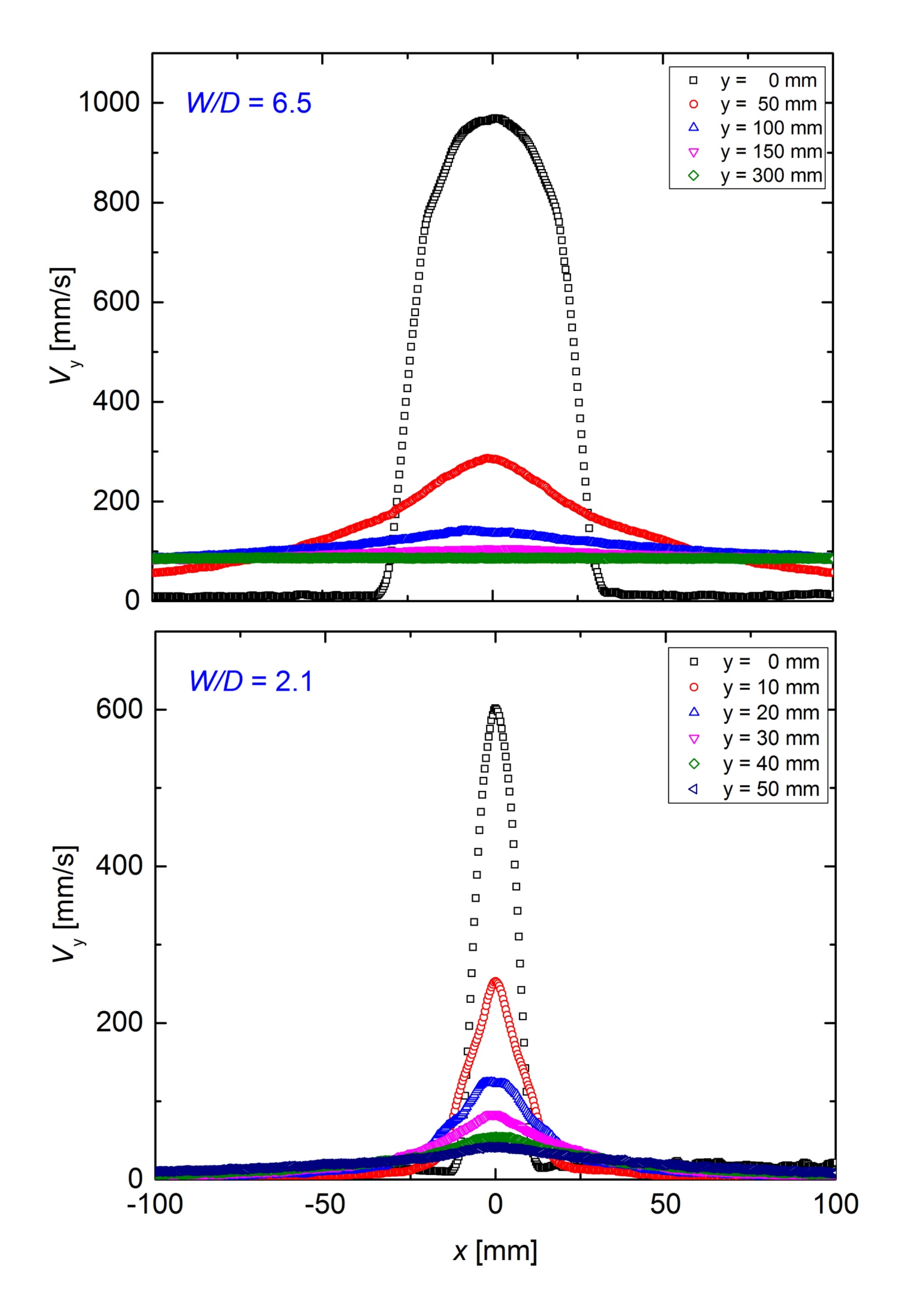}
	\caption{
Flow profiles at orifice sizes $W=60$~mm and $20$~mm. In heights from $\approx 2.5~W$ above the orifice, the vertical flow profile becomes smooth. In the upper part, we find simple plug flow.
(Only 20 cm of the full bin width are shown.)
}
\label{fig:flowprofiles}
\end{figure}

The flow field was measured by means of particle tracking, and a coarse-graining technique (comparable to Ref.~\cite{Weinhart2016})  was used to smoothen the flow profiles. It turns out that these profiles are also
essentially different from that of hard particles. Figure \ref{fig:flowfield} shows the horizontal and vertical flow components. It is evident that the flow is not restricted to a funnel-shaped region above
the outlet, but that the particles also move sidewards along the bottom of the container (Fig. \ref{fig:flowfield}a). The vertical flow component becomes rather uniform at levels approximately $3 W$
above the outlet (Fig.~\ref{fig:flowfield}b). Figure~\ref{fig:flowprofiles} shows the mean vertical flow velocity at different height levels determined from particle tracking.

The upper part of the silo performs simple plug flow.
The hexagonal lattice may be the primary reason for this feature; a study of bidisperse or polydisperse
hydrogel spheres is required to test this hypothesis.

\begin{figure}[htbp]
\includegraphics[width=0.88\columnwidth]{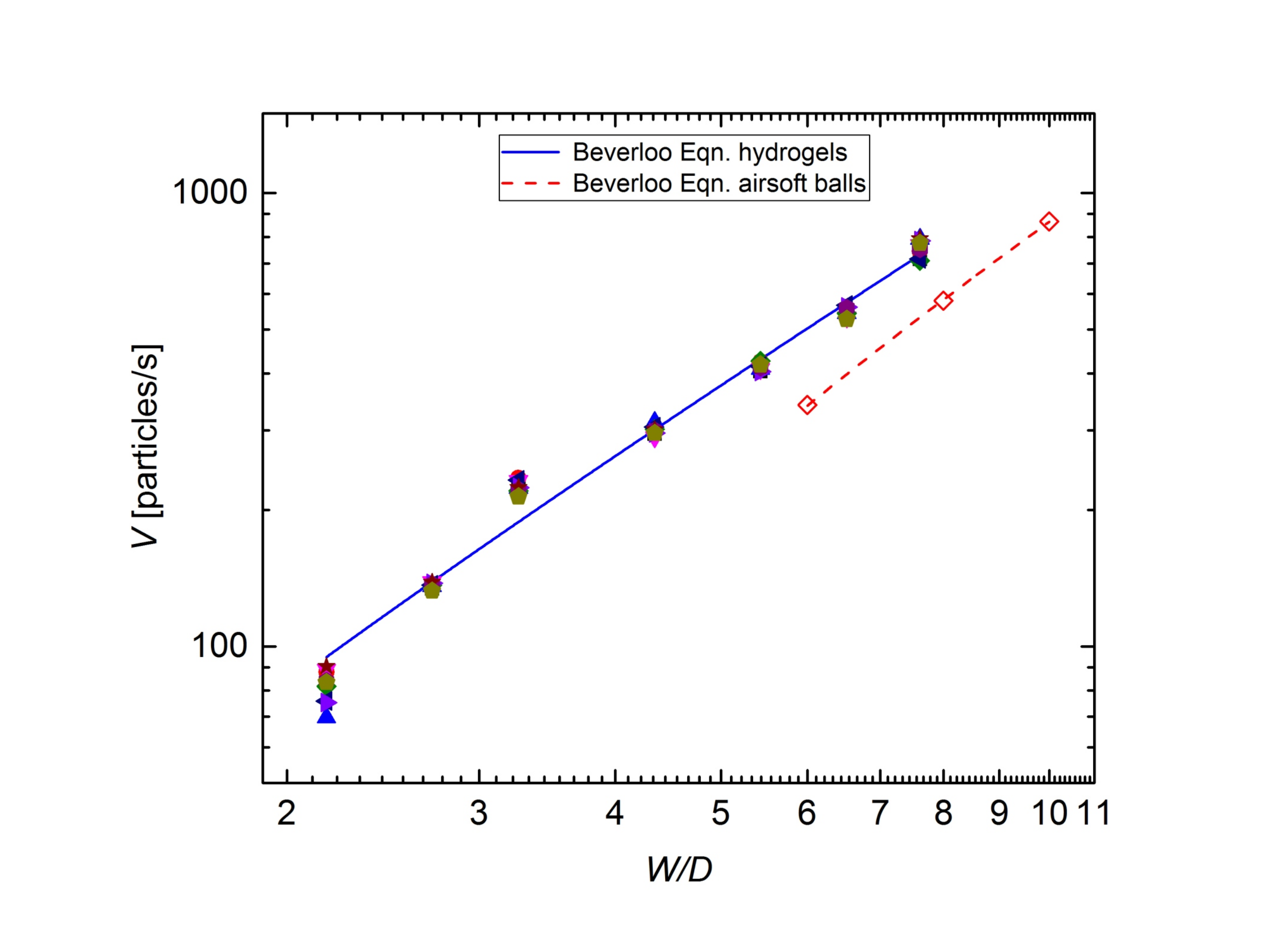}
	\caption{
Comparison of the flow rates vs. $W/D$ for soft hydrogel spheres and hard airsoft balls.
It is evident that both can be fitted with Eqn. (\ref{eq:beverloo}), but the soft particles show a substantially
higher flow rate at given relative orifice sizes, and the coefficient $k=0.3$ for the soft spheres is
extremely small.
}
\label{fig:beverloo}
\end{figure}

Figure~\ref{fig:beverloo} shows the average discharge rate $V$ in spheres per second
for different $W/D$ ratios.
The solid line is a fit with Beverloo's Eqn.
\begin{equation}
\label{eq:beverloo}
V=\dot N = V_0 (W/D-k)^{1.5}
\end{equation}
with the parameters $V_0 = 37$~ particles/s~$=15.5$~g/s and $k=0.3$. Compared to
hard particles, where the parameter $k$ is of the order of 1.5, this value is very small,
this means that the flow velocity through narrow orifices is much more effective. For hard
spheres, Thomas et al.~\cite{Thomas2015} have shown that Beverloo's equation fits the continuous
flow of hard particles at high opening sizes as well as the flow within avalanches through narrow openings, and
that there is no discontinuity in the dynamic parameters. This was confirmed in our earlier
experiments with non-spherical hard grains \cite{Ashour2017}.
For the soft spheres, the validity of Beverloo's equation cannot be tested in the limit
of small orifice sizes. The reason is that the flow rate fluctuates considerably when $W<2D$
(Fig.~\ref{fig:outflow}), and even non-permanent clogs can be observed around $W\approx 2D$
(the flow pauses for up to several seconds and restarts after a non-permanent arch above the orifice has resolved). Moreover, in our setup the influence of the fill height on the flow rate becomes non-negligible.
As is seen in Fig.~\ref{fig:outflow}, the flow rate slightly decreases even in the $W\approx 2.06$
geometry when the container is half-empty. Therefore, the lowest datum points in Fig.~\ref{fig:beverloo}
($W\approx 2.18D$) deviate systematically from the prediction.

Summarizing, the outflow of soft, practically frictionless hydrogel spheres distinguishes
qualitatively in several aspects from that of previously investigated hard particles. First, the
extremely low friction coefficient leads to hydrostatic pressure conditions in the
2D bin. Second, the softness of the particles leads to a pressure-dependence of the clogging probability at small orifice widths. The combination of these two effects results in a complex dependence
of the clogging characteristics on the fill height of the container. When $W/D$ is larger than 2.2, the silo empties
without clogging (only a few particles in a single layer remain at the horizontal container bottom).
When the ratio $W/D$ becomes smaller, the flow fluctuates, non-permanent arches may form and block
the outflow up to several seconds. Permanent arches block the outlet when the pressure at the bottom
becomes lower than a certain critical value, i. e. when the fill level of the silo reaches a critical
height. At the smallest openings where we found avalanches in our container, $W\approx 1.5 D$, the clogs
formed when the pressure at the orifice became less than about 5~kPa. We assume here that the pressure measured in a stationary state is comparable to the pressure during slow outflow. The pressure at the container bottom suffices to deform the spheres by at least 20\% of their radii. This allows the spheres to adopt their shapes under the action of force chains, to leave intermittent clogs, and to pass through narrow outlets as long as a certain pressure is maintained. For hard particles,
there is practically no influence of the fill level on the outflow unless the hopper is
emptied to a very shallow layer. The outflow is practically pressure-independent, even though the absolute pressure for comparable particle densities is much lower than for the hydrogel spheres (Fig.~\ref{fig:pressure}).

The flow profile inside the container distinguishes from that of hard spheres with friction in
that the flow involves all particles, even those in the first layer on a flat horizontal bottom. There is {\em no} stagnant zone in our bin. The downward flow in the upper part (approximately $2.5W$ above the bottom) is constant,
practically independent of horizontal and vertical positions. This is primarily due
to the monodispersity of the material, which generates a nearly perfect hexagonal packing, so that the
material far above the opening is displaced downward as a crystallized structure. We assume that this may be different
with bidisperse or polydisperse hydrogel spheres, it is not necessarily related to softness or low friction  of our material. Finally, the equation derived by Beverloo for the discharge rate can be used to satisfactorily describe the outflow as a function of the relative opening width $W/D$.
The equation becomes meaningless, however, when $W/D<2$ because the flow is neither constant nor
pressure-independent when the outlet becomes comparable to one or two particle diameters.

Hong and Weeks \cite{Weeks2017c} recently reported hopper experiments with hydrogel spheres. Their observations were in qualitative agreement with our findings. In their study, the focus was
on very small numbers of particles, so that a pressure dependence could not be tested, and the hopper geometry differed from ours in that the bottom was trapezoid-shaped.

Compared to most of the previously investigated materials, the hydrogel spheres differ in two aspects,
they are soft, and they have very low friction. In the first respect, they are between air and oil
bubbles and hard materials, much closer to hard spheres. In the second respect, they are much closer to
the fluid inclusions. In view of a general understanding of silo discharge, it will be of primary importance to discriminate the individual influences of both features on the outflow characteristics. And in addition, a study of hydrogels in 3D hoppers will be needed to allow a better comparison with realistic practical situations.

\begin{acknowledgments}
We thank Levente Kocsis for participation in the setup of the experiment and preliminary measurements.
Financial support by the NKFIH (grant No. OTKA K 116036) is acknowledged.

\end{acknowledgments}

\bibliographystyle{apsrev}

\end{document}